\documentclass[iop]{emulateapj}

\newcommand\rms[1]{\langle #1^2 \rangle^{1/2}}
\newcommand\ms[1]{\langle #1^2 \rangle}
\newcommand\ave[1]{\langle #1 \rangle}
\newcommand{\rev}[1]{#1}

\begin{document}

\title{
Importance of Giant Impact Ejecta for Orbits of Planets Formed during the Giant Impact Era}

\shorttitle{
Importance of Giant Impact Ejecta
}

%\correspondingauthor{Hiroshi Kobayashi}
\email{hkobayas@nagoya-u.jp}

\author{Hiroshi Kobayashi}
%\affiliation{Department of Physics, Nagoya University, Nagoya, Aichi 464-8602, Japan}

\author{Kazuhide Isoya}
%\affiliation{Department of Physics, Nagoya University, Nagoya, Aichi 464-8602, Japan}

\author{Yutaro Sato}
\affiliation{Department of Physics, Nagoya University, Nagoya, Aichi 464-8602, Japan}

\begin{abstract}
Terrestrial planets are believed to be formed via giant impacts of
 Mars-sized protoplanets. Planets formed via giant impacts have highly
 eccentric orbits. A swarm of planetesimals around the planets may lead
 to eccentricity damping for the planets via the equipartition of
 random energies (dynamical friction).  However, dynamical friction
 increases eccentricities of planetesimals, resulting in high
 velocity collisions between planetesimals. The collisional cascade grinds
 planetesimals to dust until dust grains are blown out due to radiation
 pressure. Therefore, the total mass of planetesimals decreases due to
 collisional fragmentation, which weakens dynamical friction. We
 investigate the orbital evolution of protoplanets in a planetesimal
 disk, taking into account collisional fragmentation of
 planetesimals. For 100\,km-sized or smaller planetesimals, dynamical
 friction is insignificant for eccentricity damping of planets because
 of collisional fragmentation.  On the other hand, giant impacts eject
 collisional fragments. Although the total mass of giant impact
 ejecta is 0.1-0.3 Earth masses, the largest impact ejecta are $\sim 1,000$\,km in size. We also investigate the orbital evolution of
 single planets with initial eccentricities 0.1 in a swarm of such giant
 impact ejecta.  Although the
 total mass of giant impact ejecta decreases by a factor of 3 in 30\,Myrs,
 eccentricities of planets are damped down to the Earth level ($\sim
 0.01$) due to interaction with giant impact ejecta. Therefore, giant
 impact ejecta play an important role for determination of
 terrestrial planet orbits.  

%\keywords{ planets and satellites:
% formation --- planets and satellites: terrestrial planets --- planets
% and satellites: dynamical evolution and stability }
\end{abstract}

\section{Introduction}

In the standard scenario for terrestrial planet formation, Mars-sized
protoplanets are formed prior to the gas depletion of the protoplanetary
disk (in several Myrs) with large orbital separations $\sim 10$
mutual Hill radii \citep[e.g.,][]{kobayashi13}. The gas depletion
triggers the long term orbital instability of protoplanets and the
collisions between protoplanets induced by the orbital instability 
result in the formation of Earth or Venus sized planets,
which is called the giant impact stage \citep{chambers98,iwasaki01}.

Most of the masses of the terrestrial planets in the Solar System is in Earth and
Venus, which have low eccentricities of 0.017 and 0.007, respectively.  The largest planets
formed in orbital simulations for giant impact stages have much greater eccentricities and inclinations
than those of Earth or Venus \citep{chambers01,kokubo06}.  Those
eccentricities and inclinations are possible to be damped via the
equipartition of random energies (dynamical friction) with surrounding
planetesimals \citep{obrien06,raymond09,morishima10}. However, the
surface density of surrounding planetesimals decreases via the collisional
cascade of the planetesimals \citep{kobayashi10}, which reduces the
efficiency of dynamical friction. Therefore, collisional fragmentation
plays an important role in this issue.

On the other hand, a series of giant impacts eject fragments with a
total mass comparable to Earth, \rev{ resulting in the increase of
infrared emission. Therefore, giant impact ejecta
explain infrared excesses caused by warm debris disks around 1AU
\citep{genda15}, while cold debris disks beyond 10\,AU may be related to
collisional fragmentation in planetesimal disks induced by planet
formation \citep{kobayashi14}.  } Such giant-impact-ejecta disks may
affect the orbital evolution of protoplanets. The evolution of total
masses of giant impact ejecta is controlled by the collisional cascade.
Therefore, we need to consider the orbital evolution and the collisional
cascade simultaneously.

The orbital evolution of protoplanets in the giant impact stage is
mainly treated by $N$-body simulation. However, all fragments produced
via collisional fragmentation cannot be treated individually by $N$ body
simulation because of computational limitation. Therefore, one applies
the super-particle approximation where a super particle represents a
large number of planetesimals and fragments.  This method is applied for
planet formation \citep{levison12,morishima15,walsh19} and for debris
disks \citep{kral13,nesvold13}.  In this paper, we newly develop the $N$
body code including the mass evolution of planetesimals via collisional
cascade, which allows us to evaluate dynamical friction and collisional
fragmentation in the giant impact stage. In \S.~\ref{sc:method}, we
explain the method to develop the code. In
\S.~\ref{sc:collisional_cascade}, we conduct test calculations for the
collisional cascade and validate the method. In \S.~\ref{sc:hybrid_sim},
we perform simulations for the orbital evolution of protoplanets in
planetesimal or giant-impact-ejecta disks using the newly developed
code. In \S.~\ref{sc:discussion}, we discuss the effect of remnant
planetesimals and giant impact ejecta on the orbital evolution of
protoplanets in the giant impact stage.  We summarize our finding in
\S.~\ref{sc:summ}.

\section{Method}
\label{sc:method}

In the giant impact stage, the orbital evolution and collisions of
protoplanets occur. The gravitational interaction with a planetesimal
disk is important for the final orbits of protoplanets. However, the
disk mass of planetesimals decreases due to the collisional cascade
starting from planetesimal fragmentation. Therefore, we need to treat
the evolution of orbits and masses consistently. 

We apply the super-particle approximation for planetesimals and smaller
bodies ejected by collisional fragmentation; a super particle represents
to planetesimals and collisional fragments. Meanwhile, we apply a single
particle for a single protoplanet.  We numerically integrate the
equations of motion of particles via the fourth order Hermite scheme
\citep{makino92,kokubo04}.  The orbital integration allows us to
accurately treat dynamical evolution and direct collisions between
protoplanets and super particles\footnote{\rev{ The total ejecta mass
caused by collisions between protoplanets and super particles is
estimated to be much smaller than the disk mass and their collisional
lifetimes are much shorter than that of the disk. Therefore, we ignore
collisional ejecta caused by collisions between protoplanets and super
particles.}}.  However, the number of super particles that we apply is
much smaller than that of planetesimals and fragments with which we are
concerned so that statistical treatment is required for accurate
calculation of interactions between super particles.

\citet{morishima15} developed a method for 
the collisions and dynamical interactions
between super particles in $N$-body simulations
We treat the collisions between super particles following \citet{morishima15}, while we ignore the dynamical interaction because the
collisional timescale is much shorter than the
dynamical interaction timescale for dynamically hot planetesimals. 

We consider the $j$-th super particle at the cylindrical coordinate
($r_j,\theta_j,z_j$) originated at the host star with mass $M_*$.  The
surface density around the super particle, $\Sigma_j$, is determined by
the total mass of super particles in the area with $r = [r_j - \delta
r_j :r_j + \delta r_j]$ and $\theta = [\theta_j - \delta \theta_j:
\theta_j + \delta \theta_j]$;
\begin{equation}
 \Sigma_j = \frac{1}{4 r_j \delta r_j \delta \theta_j}
  \left[m_j + \sum_k^{N_{{\rm n},j}} m_k \right],\label{eq:sigma}
\end{equation}
where $N_{{\rm n},j}$ is the number of super particles in the area and $m_k$ is
the mass of the $k$-th super particle in the area. We choose $\delta r$ and $\delta
\theta$ according to the accuracy of mass evolution of planetesimals due
to the collisional cascade, which we discuss in \S
\ref{sc:collisional_cascade}.

The relative velocity, which characterises collisional fragmentation of
planetesimals in super particles, is determined by the orbital elements 
of super particles in the area. We calculate the relative velocity $v_j$ between planetesimals
in the $j$-th super particle with semimajor axis $a_j$, eccentricity
$e_j$, inclination $i_j$, the longitude of pericenter $\varpi_j$, and
the longitude of ascending node $\Omega_j$, given by 
\begin{equation}
 v_{{\rm r},j} = v_{\rm K} \sqrt{e_{{\rm r},j}^2+i_{{\rm r},j}^2},\label{eq:vel} 
\end{equation}
where $v_{K} = \sqrt{G M_* /a_j}$, $G$ is the gravitational constant,
and 
\begin{eqnarray}
 e_{{\rm r},j}^2 &=& \frac{1}{N_{{\rm n},j}} \sum_k^{N_{{\rm n},j}} [e_j^2 + e_k^2 -2 e_j e_k
  \cos(\varpi_j-\varpi_k)],\label{eq:rel_e} \\
 i_{{\rm r},j}^2 &=& \frac{1}{N_{{\rm n},j}} \sum_k^{N_{{\rm n},j}} [i_j^2 + i_k^2 -2 i_j i_k
  \cos(\Omega_j-\Omega_k)].\label{eq:rel_i} 
\end{eqnarray}

The collisional cascade grinds planetesimals down to micron-sized grains,
which are blown out due to the radiation pressure of the host star \citep[e.g.,][]{kobayashi08,kobayashi09,krivov10}. Collisional
fragmentation of planetesimals thus reduces the surface density of
planetesimals due to the collisional cascade. 
We consider the quasi-steady-state collisional cascade, for which 
\citet{kobayashi10}
derived the reduction rate of the surface density of planetesimals, 
given by 
\begin{equation}
 \frac{d\Sigma}{dt} = - \frac{(2-\alpha_c)^2}{m_{\rm c}^{1/3}} \Sigma^2
  \Omega_{\rm K} \left(\frac{v_{\rm r}(m_{\rm c})^2}{2 Q_{\rm
		  D}(m_{\rm c})^*}\right)^{\alpha_c-1} f(\alpha_c),\label{eq:dsigma_dt} 
\end{equation}
where $m_{\rm c}$ is the mass of largest bodies in the collisional cascade,
$\alpha_c$ is the index of the mass distribution of bodies determined by 
the collisional cascade, $f(\alpha_c)$ is the dimensionless value dependent on
$\alpha_c$, and $Q_{\rm D}^*$ is the specific impact energy needed for the ejection of
the half mass of colliders. 
\rev{The mass of largest bodies $m_{\rm c}$ does not change in the
collisional cascade, which is valid until the bodies with
$m_{\rm c}$ exist.}
For $v_{\rm r}^2/ Q_{\rm D}^* \propto m^p$,
$\alpha_c$ is given by \citep{kobayashi10}
\begin{equation}
 \alpha_c = \frac{11+3p}{6+3p}.\label{eq:alpha_c} 
\end{equation}
\rev{
The mass distribution with the index $\alpha_{\rm c}$ is achieved in the
timescale for the collisional cascade. 
An earlier mass distribution index 
may be different from $\alpha_{\rm c}$ in Eq.~(\ref{eq:alpha_c}) . 
However, the mass loss due to the 
collisional cascade is negligible in such an early stage 
and the mass evolution mainly
occurs when the mass distribution given by Eq.~(\ref{eq:alpha_c}) is
achieved. 
}
Therefore, 
Eqs.~(\ref{eq:dsigma_dt}) and (\ref{eq:alpha_c}) are valid to
treat the surface density evolution accurately. 
According to \citet{kobayashi10}, $f(\alpha_c)$ is given by 
\begin{equation}
 f(\alpha_c) = 1.055 \rho^{-2/3} %h_0 
  \left[ \left( - \ln \epsilon + \frac{1}{2-b}\right) s_1
		 (\alpha_c) + s_2(\alpha_c) + s_3(\alpha_c)\right], 
\end{equation}
where $\rho$ is the density of bodies, 
$b$ is the power-law index of the mass distribution of fragments, 
$\epsilon$ is the constant determining the largest mass of fragments,
and\footnote{\rev{
We correct a typo in $s_3(\alpha_{\rm c})$ of \citet{kobayashi10}.}} 
\begin{eqnarray}
 %h_0 &=& , \\
 s_1(\alpha_c) &=& \int_0^\infty d \phi \frac{\phi^{1-\alpha_c}}{1+\phi}, \\
 s_2 (\alpha_c) &=& - \int_0^\infty d \phi \frac{\phi^{1-\alpha_c}}{1+\phi}
  \ln \frac{\phi}{(1+\phi)^2}, \\
 s_3 (\alpha_c) &=& \int_0^\infty d \phi \frac{\phi^{-\alpha_c}}{1+\phi}
  \ln (1+\phi).
\end{eqnarray}
The impact laboratory experiments shows $b = 1.5-1.7$ and $\epsilon \sim
0.1$ \citep[e.g.,][]{takagi84,nakamura91} so that the choice of $b$ and $\epsilon$
insignificantly change the value of $f(\alpha_c)$.\footnote{
\rev{We here
evaluate the dependence of $f(\alpha_{\rm c})$ on $b$. 
We consider
$p = 0.453$, and then $\alpha_c = 1.68$. The values of $s_1$, $s_2$, and
$s_3$ are calculated to be $3.7, 7.4$, and $2.6$, respectively. The
function $f(1.68)$ with $\epsilon = 0.2$ is estimated to be 28.9 and 33.8 for $b = 1.5$ and 1.7, respectively. }
} Therefore, we set $b =
3/5$ and $\epsilon = 0.2$. 

If $v_{\rm r}$ is fixed, the integration of Eq.~(\ref{eq:dsigma_dt}) over time $t$
gives 
\begin{equation}
 \Sigma(t) = \frac{\Sigma(0)}{1 + t/\tau_0},\label{eq:sigma(t)} 
\end{equation}
where $\tau_0 = - \Sigma(0)/\dot \Sigma(0)$. In the giant impact stage,
Eq.~(\ref{eq:sigma(t)}) is not always valid because of the evolution of
$v_{\rm r}$. Therefore, we use
Eq.~(\ref{eq:sigma(t)}) only for the validation of our simulation in \S \ref{sc:collisional_cascade}. 

%\begin{figure}
% \plotone{cascade_mass_evo.pdf}
% \figcaption{Surface density evolution of a planetesimal disk via
% collisional cascade. The red solid curve indicate the result of the
% simulation, while the black dashed curve is given by  the analytical solution. 
% \label{fig:collision_cascade}
%}
%\end{figure}

Based on Eq. (\ref{eq:dsigma_dt}), the mass-loss rate of the $j$-th
super particle due to the collisional cascade is given by 
\begin{equation}
\frac{1}{m_j} \frac{d m_j}{dt} = 
 - \frac{(2-\alpha_c)^2}{m_{{\rm c},j}^{1/3}} 
 \Sigma_j \Omega_{\rm K} \left(\frac{v_{{\rm r},j}^2}{2 Q_{\rm
		  D}(m_{{\rm c},j})^*}\right)^{\alpha_c-1} f(\alpha_c),\label{eq:mass_loss_particle}  
\end{equation}
where $m_{{\rm c},j}$ is the mass of the largest bodies in the $j$-th
super particle. 
We calculate the mass evolution of super particles via the integration
of Eq. (\ref{eq:mass_loss_particle}) 
using Eqs.(\ref{eq:sigma}) and (\ref{eq:vel}). 

The masses of the largest bodies in super particles, $m_{\rm c}$, are set to
$\ga 10^{16}\,$g corresponding to $\ga 1$\,km in radius. Therefore,
$Q_{\rm D}^*$ of such a body is mainly determined by shuttering and
gravitational reaccumulation
\citep{benz99,leinhardt12,jutzi15,genda15b,genda17,suetsugu18}. 
Therefore, $Q_{\rm D}^*$ has a monotonous increasing function of $m_{\rm
c}$, which is assumed to be \citep[e.g.,][]{benz99}
\begin{equation}
 Q_{\rm D}^* = Q_{\rm 0} \left( \frac{m_{\rm c}}{10^{21}\,{\rm g}} \right)^{p}, 
\end{equation}
where $Q_{\rm 0}$ and $p$ are determined from numerical simulations. 
The collisional cascade is controlled not by individual collisions but by
successive collisions. The value averaged over impact angles is applied
for $Q_{\rm D}^*$. Taking into account self-gravity and a model of rock
fractures, \citet{benz99} obtained the averaged values for $Q_0$ and $p$
via the smoothed particle hydrodynamics (SPH) simulation. 
Recently, the dependence of 
$Q_{\rm D}^*$ on the number of SPH particles is argued and the
value of $Q_{\rm D}^*$ decreases 20\% in the limit of high resolution
simulation \citep{genda15}. However, the friction of damaged rock, which
was not considered in \citet{benz99}, increases $Q_{\rm D}^*$
\citep{jutzi15}. 
The value of $Q_{\rm D}^*$ obtained in \citet{benz99} is similar to that
given by the high-resolution simulations with the fraction
\citep{suetsugu18}. 
Therefore, we set $Q_0 = 9.5 \times 10^8\,$erg/g and $p
= 0.453$ \citep{benz99}. 

\section{Validation for Collisional Cascade}
\label{sc:collisional_cascade}

We perform a simulation for collisional evolution of a planetesimal disk
composed of 2,000 super particles with $a = 0.975$--1.025\,AU and $m_{\rm c} = 10^{16}\,$g. 
The radial distribution of super particles is put according to
$\Sigma(a) \propto a^{-1}$ and $e$ and $i$ have the Rayleigh distributions 
with mode values $e = 0.01 (a/{\rm 1 \, AU})^{1/2}$ and $i = 0.005
(a/{\rm 1 \, AU})^{1/2}$. 
\rev{
The collisional-cascade timescale in this setting 
is estimated to be $\tau_0 \approx
0.6$ years. We evaluate the accuracy of the method from the comparison
with the analytic solution in Eq.~(\ref{eq:sigma(t)}) at $t=100$ years. 
}

%Fig.~\ref{fig:collision_cascade} shows the 
%surface density of planetesimals averaged 
%over the disk. We set $\delta r = 0.01 \,{\rm AU}$ and $\delta \theta
% = \pi$. The collisional cascade decreased the surface
%density of planetesimals. The result of the simulation reproduces the
%analytical solution given by Eq.~(\ref{eq:sigma(t)}). 

\begin{figure}
 \plotone{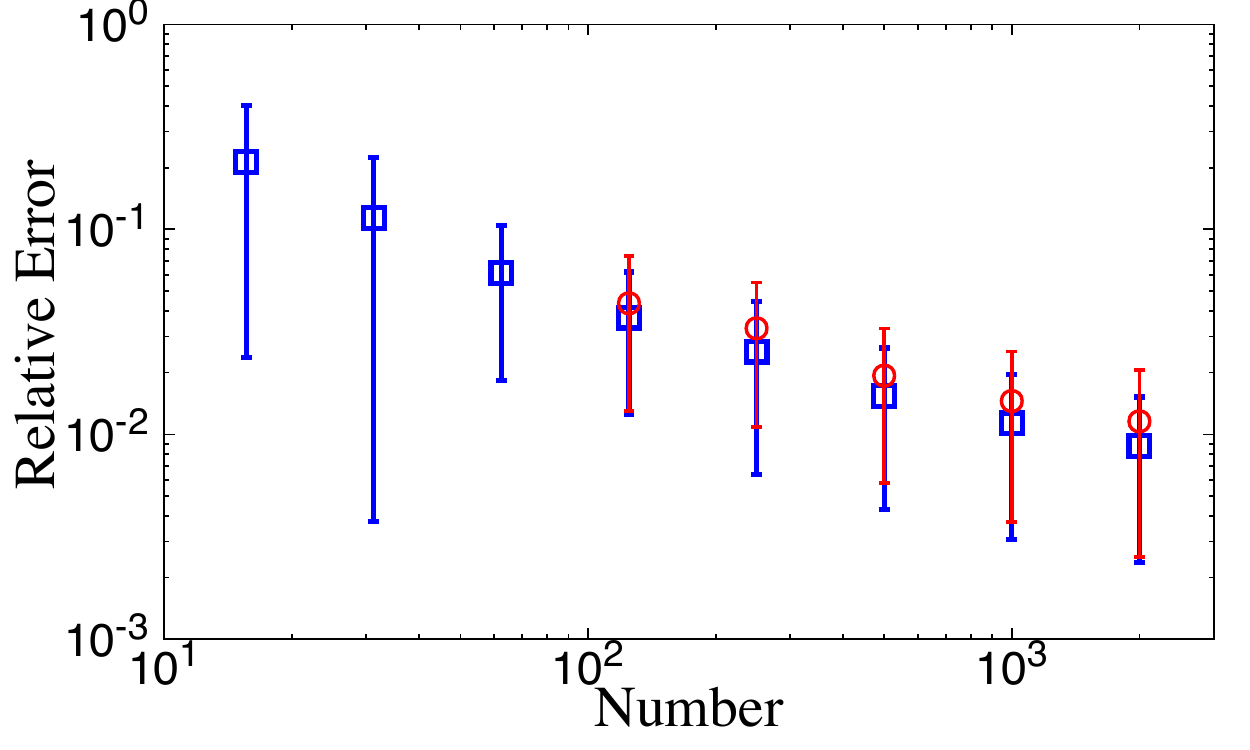}
 \figcaption{
 Relative errors of the surface density at $t=100$ years with the
 setting of $\tau_0 \approx 
 0.6$ years for $\delta \theta
 = \pi/8$ (red circles) and $\pi$ (blue squares), as a function
 of the number of super particles. 
 The error bars indicate the standard errors obtained from 40 runs. 
 }
\label{fig:comp_cc}
\end{figure}

The accuracy of the simulation depends on the number of particles in the
neighbor area, $N_{{\rm n}}$.  We calculate the relative error of the
surface density from comparison of the simulation with the analytical
solution.  Fig.~\ref{fig:comp_cc} shows the dependence of relative
errors on the number of super particles.  The errors are roughly
proportional to the number of super particles, which is proportional to
$N_{\rm n}$ for fixed $\delta \theta$.  However, the errors are almost
independent of $\delta \theta$. Although $N_{\rm n}$ decreases with decreasing $\delta \theta$, the number of particles having the experience
in a neighbor area is almost same in the time much longer than orbital
periods if the number of super particles and $\delta r$ are fixed. The
timescale of the collisional cascade is mainly much longer than orbital
periods. Therefore, the accuracy better than 10\% is obtained if the
number of particles in the
annulus with width $\delta r$ is $\ga 10$. 

\section{Orbital Evolution of Protoplanets in a Swarm of Fragmenting Planetesimals}
\label{sc:hybrid_sim}

We carry out simulations for the orbital evolution of 3 protoplanets
with an Earth mass $M_\oplus$ in a planetesimal disk with $30 M_\oplus$
composed of 3,000 super particles around the central star with solar
mass $M_\odot$.  The semimajor axis of the intermediate protoplanet is
initially set at 1\,AU, the orbital separation of protoplanets is 10
mutual Hill radii, and their eccentricities and inclinations are 0.03
and 0.015, respectively.  The radial distribution of super particles is
initially put according to $\Sigma(a) \propto a^{-1}$ and their $e$ and
$i$ have the Rayleigh distributions with mode values $e = 0.03$ and $i =
0.015$, respectively.  The intermediate radius and width of the
planetesimal disk are 1\,AU and 30 mutual Hill radii, respectively. 
For the treatment of collisional fragmentation, we put $\delta r = 0.01$\,AU and $\delta \theta = \pi/8$. 

\begin{figure}[htb]
 \plotone{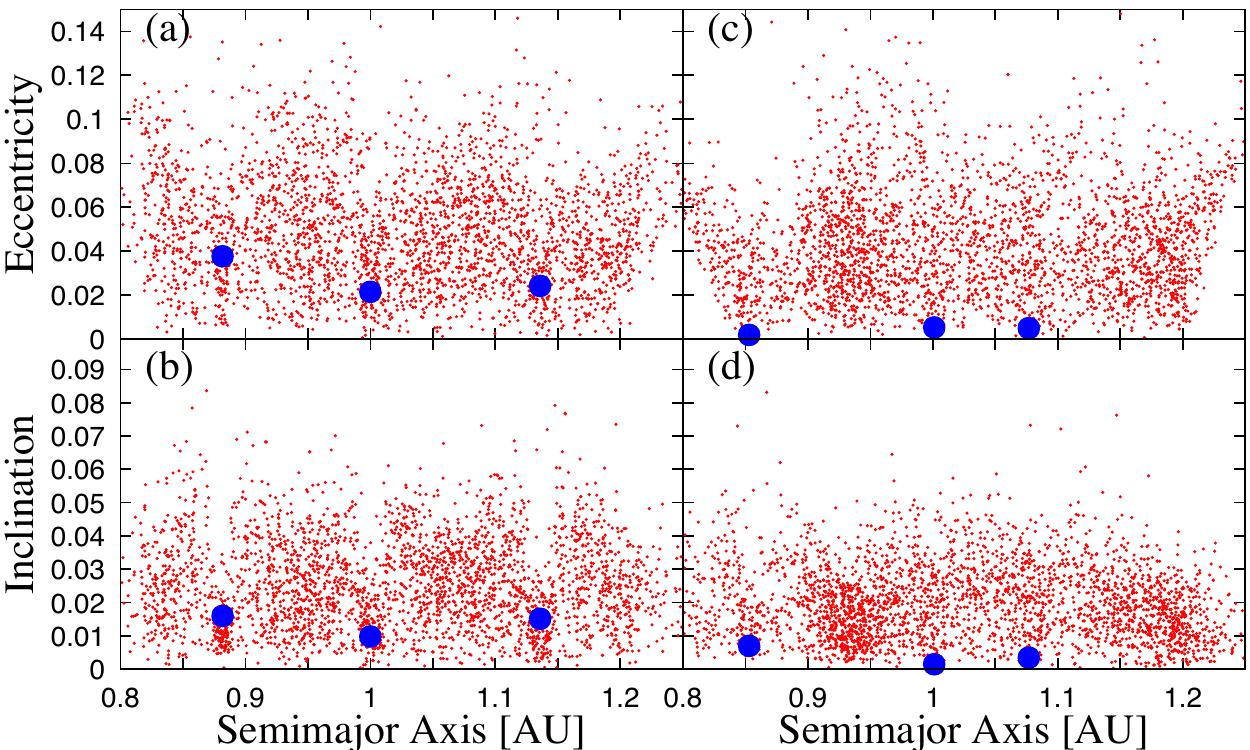} \figcaption{ Eccentricities (a,c) and
 inclinations in radian (b,d) of super particles (red dots) and
 protoplanets (blue fulled circles) with fragmentation (a,b) or without
 fragmentation (c,d).  } \label{fig:orbital_dist}
\end{figure}

The collisional cascade is characterized by planetesimal mass $m_{\rm
c}$. We set $m_{\rm c} = 10^{\rm 16}$\,g ($\approx 1\,$km in radius) for
collisional fragmentation, while we have an additional simulation
without collisional fragmentation ($Q_{\rm D}^* = \infty$ or $m_{\rm c}
= \infty$).  Figure~\ref{fig:orbital_dist}a,b shows the orbital
distribution of protoplanets and super particles at $t = 10^3$\,years
with collisional fragmentation, while Figure~\ref{fig:orbital_dist}c,d
is that without fragmentation.  Dynamical friction decreases $e$ and
$i$ of protoplanets, while $e$ and $i$ of planetesimals
increase. Increases in $e$ and $i$ of planetesimals activate their
collisional cascade, which reduces the surface density of
planetesimals. Collisional fragmentation weakens dynamical friction so
that $e$ and $i$ of protoplanets with fragmentation remain higher than
those without fragmentation.

\begin{figure}[htb]
 \plotone{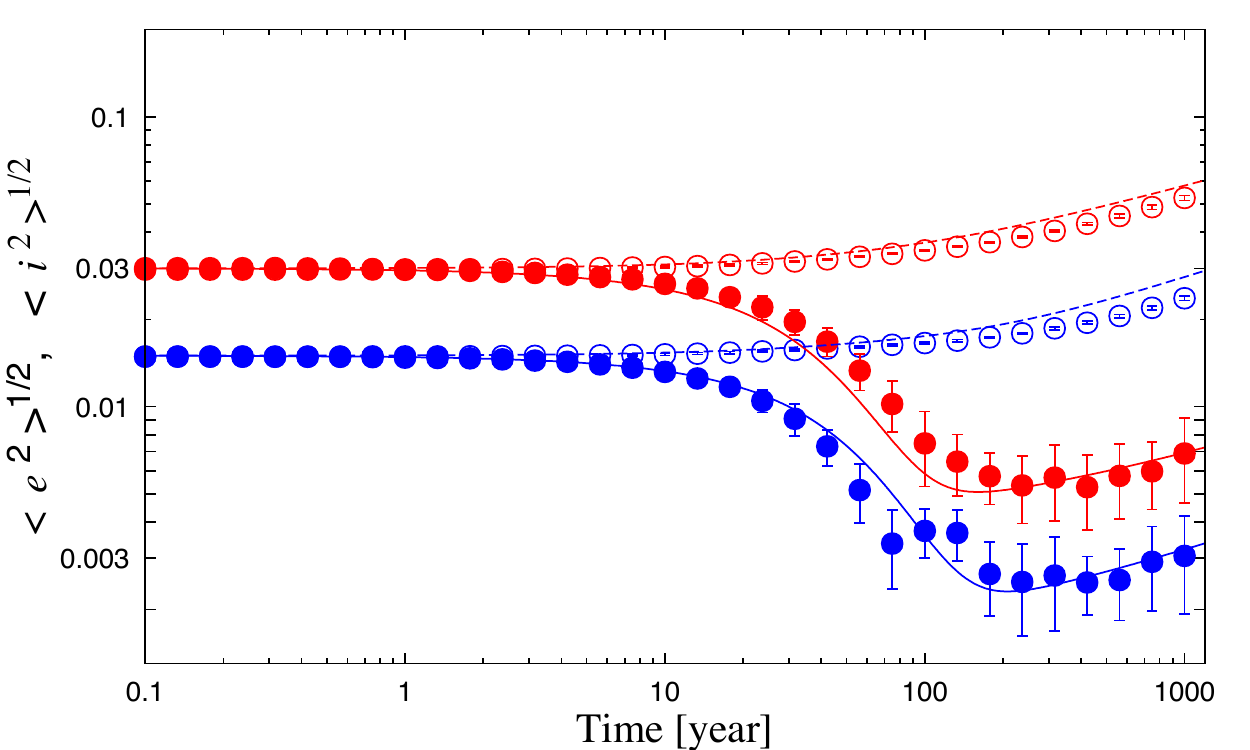} \figcaption{The root mean squares of
 eccentricities $\rms{e}$ (red) and inclinations $\rms{i}$ (blue) for
 planetesimals (open circles) and protoplanets (filled circles) without
 collisional fragmentation in the planetesimal disk same as
 Fig.~\ref{fig:orbital_dist}. The error bars are given by the standard
 deviation of 13 runs. Analytic solutions $\rms{e}$ and $\rms{i}$ for
 planetesimals (dotted lines) and protoplanets (solid lines) is obtained
 from the time integration of Eqs. (\ref{eq:dedt}) and
 (\ref{eq:didt}). } \label{fig:evo_ecc_inc_nofrag}
\end{figure}

\begin{figure}[htb]
 \plotone{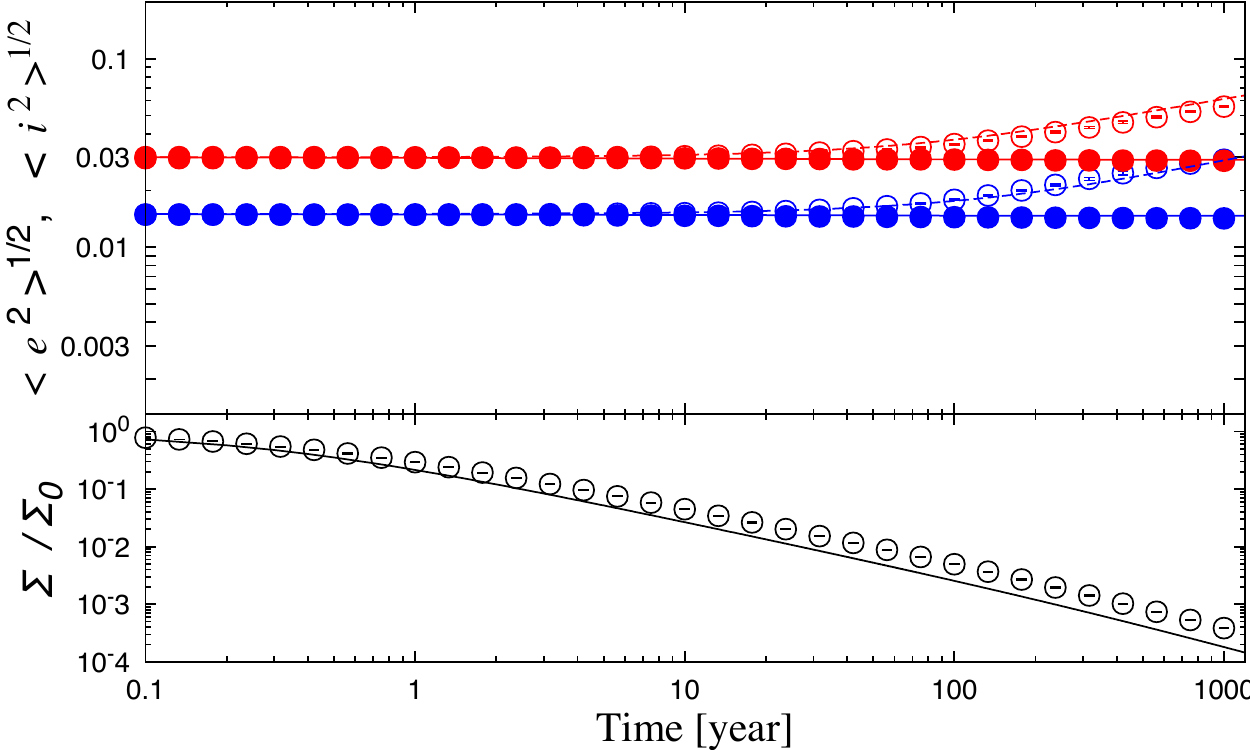}
 \figcaption{Same as Fig.~\ref{fig:evo_ecc_inc_nofrag} but
 for the treatment of collisional fragmentation. 
 We take into account collisional fragmentation for $m_{\rm c} =
 1\times10^{16}$\,{\rm g}. 
The mean surface density of planetesimals, $\Sigma$, decreases due to
 collisional fragmentation. The ratio of $\Sigma$ to initial surface
 density $\Sigma_0$ obtained from the simulation (open circles) and the
 analytical solution (solid line) is shown in the bottle panel. 
 }
\label{fig:evo_ecc_inc_frag_mc5e-18_ecc0.03}
\end{figure}

\begin{figure}[htb]
 \plotone{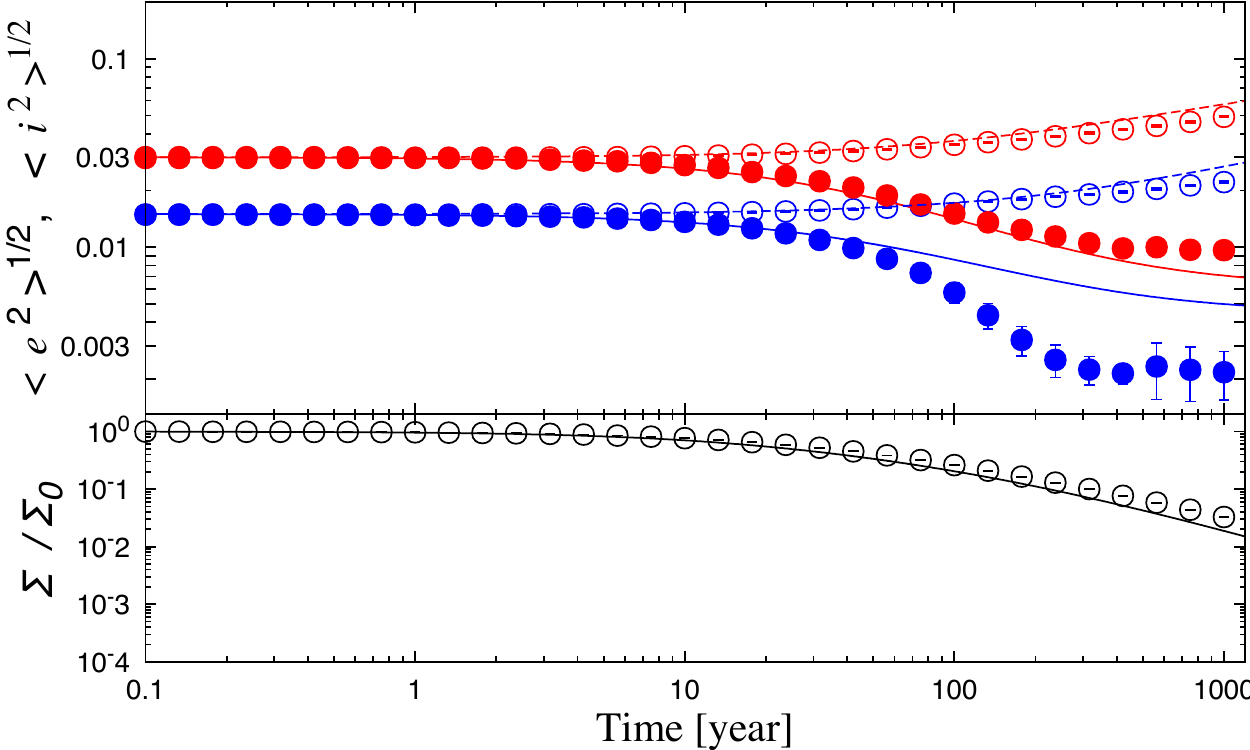}
 \figcaption{Same as Fig.~\ref{fig:evo_ecc_inc_frag_mc5e-18_ecc0.03} but
 for $m_{\rm c} = 1 \times 10^{19}$\,g. 
 }
\label{fig:evo_ecc_inc_frag_mc5e-15_ecc0.03}
\end{figure}

Figures
\ref{fig:evo_ecc_inc_nofrag}--\ref{fig:evo_ecc_inc_frag_mc5e-15_ecc0.03}
show the evolution of the root mean squares of eccentricities $\rms{e}$
and inclinations $\rms{i}$ for planetesimals and protoplanets in the
same initial setting of planetesimal disks and protoplanets as
Figure~\ref{fig:orbital_dist}.  If we ignore collisional fragmentation,
$\rms{e}$ and $\rms{i}$ of planetesimals become much larger than those
of protoplanets (Figure~\ref{fig:evo_ecc_inc_nofrag}). This is caused by
dynamical friction. However, collisional fragmentation of planetesimals
decreases the surface density of planetesimals
(Figures.~\ref{fig:evo_ecc_inc_frag_mc5e-18_ecc0.03} and
\ref{fig:evo_ecc_inc_frag_mc5e-15_ecc0.03}), which suppresses the $e$
and $i$ reduction of protoplanets via dynamical friction.  Collisional
fragmentation is effective for small $m_{\rm c}$ so that $\rms{e}$ and
$\rms{i}$ of protoplanets insignificantly change.  On the other hand,
the $e$ and $i$ evolution for planetesimals are almost independent of
fragmentation. This is because the viscous stirring of protoplanets
controls $e$ and $i$ of planetesimals.

If we consider two populations of bodies such as protoplanets 
with $\rms{e}_1$, $\rms{i}_1$, and 
mass $m_{{\rm p},1}$ and planetesimals with $\rms{e}_2$, $\rms{i}_2$,
and  mass $m_{{\rm p},2}$, 
the time differential of $\rms{e}_\alpha$ and $\rms{i}_\alpha$ for
$\alpha = 1$ or 2 (protoplanets or planetesimals) due to dynamical friction and viscous
stirring is analytically given by \citep{ohtsuki02}
\begin{eqnarray}
 \frac{d \ms{e}_\alpha}{dt} &=& a_0^2 \Omega \sum_{\beta=1,2}
  \left[
   N_{s,\beta} \frac{h_{\alpha,\beta}^4 m_{{\rm p},\beta}}{(m_{{\rm
   p},\alpha} + m_{{\rm p},\beta})^2}
   \biggl(
m_{{\rm p},\beta} P_{\rm VS}
\right.
%\right. 
\nonumber \\
& & 
\left.
+ \frac{m_{{\rm p},\beta}
    \ms{e}_\beta-m_{{\rm p},\alpha}
    \ms{e}_\alpha}{\ms{e}_\alpha+\ms{e}_\beta} P_{\rm DF}\biggr)
  \right],
\label{eq:dedt}
\\
 \frac{d \ms{i}_\alpha}{dt} &=& a_0^2 \Omega \sum_{\beta =1,2}
  \left[
   N_{s,\beta} \frac{h_{\alpha,\beta}^4 m_\beta}{(m_{{\rm
   p},\alpha} + m_{{\rm p},\beta})^2}
   \biggl(m_{{\rm p},\beta} Q_{\rm VS} 
\right.
\nonumber \\
& & 
\left.
%\left.
    + \frac{m_{{\rm p},\beta} \ms{i}_\beta
    -m_{{\rm p},\alpha} \ms{i}_\alpha
    }{\ms{i}_\alpha+\ms{i}_\beta } Q_{\rm DF}\biggr)
  \right],
\label{eq:didt}
\end{eqnarray}
where $a_0$ is the mean semimajor axis of bodies, 
$N_{{\rm s},\beta}$ is the surface number density of bodies for $\beta =
1$ or 2, $h_{\alpha,\beta} = [(m_{{\rm p},\alpha} + m_{{\rm p},\beta})/3 M_*]^{1/3}$
is the reduced Hill radius of bodies with masses $m_{{\rm p},\alpha}$ and
$m_{{\rm p},\beta}$, $P_{\rm VS}$, $P_{\rm DF}$, $Q_{\rm VS}$, and $Q_{\rm DF}$
are the efficiencies for viscous stirring and dynamical friction for $\ms{e}_{1,2}$
and $\ms{i}_{1,2}$, respectively. The analytical formulae of $P_{\rm VS}$, $P_{\rm DF}$, $Q_{\rm VS}$, and
$Q_{\rm DF}$ are given as a function of $\ms{e}_{1,2}$ and $\ms{i}_{1,2}$ in
\citet{ohtsuki02}. 

The $e$ and $i$ variation rates 
due to dynamical friction are given by the second terms on
the right hand sides of Eqs. (\ref{eq:dedt}) and (\ref{eq:didt}),
respectively. 
Protoplanets are much more massive than planetesimals ($m_{{\rm p},1}
\gg m_{{\rm p},2}$). 
For $m_{{\rm p},1} \ms{e}_1 \gg m_{{\rm p},2} \ms{e}_2$ and $m_{{\rm p},1} \ms{i}_1 \gg m_{{\rm p},2} \ms{i}_2$, 
the dynamical-friction damping rates for protoplanets is approximated to
be 
\begin{equation}
 \frac{1}{P_{\rm DF}} 
  \frac{d \ms{e}_1}{dt} \sim 
  \frac{1}{Q_{\rm DF}} 
  \frac{d \ms{i}_1}{dt} \sim 
  - \frac{a_o^2 \Omega \Sigma}{6M_*}
  \left(\frac{m_{{\rm p},1}}{3 M_*}\right)^{1/3}, 
\end{equation}
where $\Sigma$ is the surface density of planetesimals and we assume
$\ms{e}_1 \approx \ms{e}_2$ and $\ms{i}_1 \approx \ms{e}_2$. 
The damping rates are mainly determined by $\Sigma$ 
but are almost independent of the choice of planetesimal masses, 
$m_{{\rm p},2}$. This means the super-particle approximation is valid. 
However, dynamical friction leads to the equipartition of the random
energies such as $m_{{\rm p},1} \rms{e}_1 \sim
m_{{\rm p},2} \rms{e}_2$ and $m_{{\rm p},1} \rms{i}_1 \sim
m_{{\rm p},2} \rms{i}_2$. The values for the equipartition depend on
the choice of $m_{{\rm p},2}$. Therefore, we need to care the choice of
super-particle masses if we are interested in the equipartition. 
 
To compare the results of simulations with Eqs.~(\ref{eq:dedt}) and
(\ref{eq:didt}), we set $m_{{\rm p},2}$ to be super-particle mass $m_j$ and 
integrate Eqs.~(\ref{eq:dedt}) and (\ref{eq:didt}) over time (see
Fig.~\ref{fig:evo_ecc_inc_nofrag}). The analytic solution is consistent
with the simulation. For $t \ga 100$\,years, dynamical friction is
ineffective because of the achievement of energy equipartition such as 
$m_{{\rm p},1} \rms{e}_1 \sim
m_{{\rm p},2} \rms{e}_2$ and $m_{{\rm p},1} \rms{i}_1 \sim
m_{{\rm p},2} \rms{i}_2$. The result depends on the super-particle
mass as discussed above. 
However, collisional fragmentation
mainly occur prior to the achievement of the equipartition (see
Figs.~\ref{fig:evo_ecc_inc_frag_mc5e-18_ecc0.03} and
\ref{fig:evo_ecc_inc_frag_mc5e-15_ecc0.03}). 

The mass of planetesimals $m_{{\rm p},2}$ in Eqs.~(\ref{eq:dedt}) and
(\ref{eq:didt}) is initially set to be super-particle masses $m_j$,
although $m_{\rm c}$ in Eq.~(\ref{eq:dsigma_dt}) is independently given.
We then integrate 
Eqs.~(\ref{eq:dsigma_dt}), (\ref{eq:dedt}), and
(\ref{eq:didt}) over time. 
The analytic solution is roughly in agreement with the results of
simulations (see
Figs.~\ref{fig:evo_ecc_inc_frag_mc5e-18_ecc0.03} and
\ref{fig:evo_ecc_inc_frag_mc5e-15_ecc0.03}). 
In the simulations, each super particle has an independent
mass. 
%The masses of planetesimals are given by a single value $m_{{\rm p},2}$ 
%in the analytic solution, while 
Super particles in the inner disk effectively lose their masses
via collisional fragmentation because of high collisional speeds, while
super particles tend to have large masses in the outer disk. 
On the other hand, the mass evolution of planetesimal disks is
calculated with the averaged collisional speed in the analytical solution. 
Therefore, the analytic solution cannot perfectly reproduce the
simulations. However, the tendency of orbital interaction and
collisional fragmentation is understood from the analytical
solution. 

\rev{
It should be noted that the relative increases in protoplanet masses 
are smaller than 0.2 and 0.02 in the case without
and with collisional fragmentation, respectively. 
Those are caused by collisions with planetesimals. 
We estimate the effect
of collisional damping due to the collisional accretion according to the model by \citet{kobayashi16},
resulting in the $e$ damping of 0.006 and 0.0001 for protoplanets
without and with collisional fragmentation, respectively. 
The collisional damping insignificantly affects the orbital evolution of
protoplanets. Therefore, the analytical solution without collisional
damping is in good agreement with the results of simulations. 
}

The timescale of dynamical friction is estimated from Eqs.~(\ref{eq:dedt})
and (\ref{eq:didt}). If $m_{{\rm p},1} \rms{e}_1 \gg m_{{\rm p},2}
\rms{e}_2$ and $\rms{e}_2 \gg \rms{e}_1$ for $m_{{\rm p},1} \gg m_{{\rm p},2}$, 
the $\rms{e}_1$ damping timescale due to dynamical
friction has the relation as \citep{ohtsuki02}
\begin{eqnarray}
 \tau_{\rm df} &\approx& 1 \times 10^4
  \left(\frac{M_{\rm disk}}{M_\oplus}\right)^{-1}
  \left( \frac{\rms{e}_2}{0.1}\right)^2 
  \left(\frac{P_{\rm DF}}{17}\right)^{-1}
\nonumber\label{eq:tdf}
\\
&& \times 
  \left(\frac{\Delta a}{10 r_{\rm H}}\right)
  \left(\frac{a_0}{1\,{\rm AU}}\right)^{3/2} 
  \left(\frac{M_*}{M_\sun}\right)^{1/2}
\,{\rm yr}, 
%\propto \Sigma^{-1} a_0^{-1} \Omega^{-1} m_{{\rm p},1}^{-1/3} \ave{e_1^2} P_{\rm DF}^{-1}. 
\end{eqnarray}
where $M_{\rm disk}$ is the total mass of a planetesimal disk, $\Delta
a$ is the width of the disk, and 
$r_{\rm H} = (m_{{\rm p},1}/3 M_*)^{1/3} a_0$ is the Hill radius for
protoplanets. Note that $P_{\rm DF}$ is estimated to be $\sim 10$ 
for $m_{\rm p,1} \sim M_\oplus$
and $\rms{e}_1 \approx \rms{e}_2 \sim 0.1$. This estimate implies 
a planetesimal disk with $M_{\rm disk} \ll M_{\oplus}$ leads to 
eccentricity damping for Earth-sized protoplanets in a long timescale
$\gg 10^4$yr via dynamical friction. 
In addition, dynamical friction increases $\rms{e}$ of
planetesimals. The increase in $\rms{e}_2$ makes $\tau_{\rm df}$
longer. 

On the other hand, the decreasing timescale for $M_{\rm disk}$ 
due to the collisional cascade, $\tau_{\rm cc}$, 
is estimated from Eq.~(\ref{eq:dsigma_dt}) to be \citep{kobayashi10}
\begin{eqnarray}
 \tau_{\rm cc} &\approx& 4 \times 10^3 
    \left(\frac{M_{\rm disk}}{M_\oplus}\right)^{-1}
    \left( \frac{\rms{e}_2}{0.1}\right)^{-1.36}
    \left( \frac{Q_0}{9.5\times 10^8 {\rm erg/g}}\right)^{0.68}
\nonumber
\\
&& \times 
  \left(\frac{m_{\rm c}}{10^{22}\,{\rm g}}\right)^{0.79}
  \left(\frac{\Delta a}{0.1 a_0}\right)
  \left(\frac{a_0}{1\,{\rm AU}}\right)^{4.18}   
\,{\rm yr}. 
%\propto \Sigma^{-1} \Omega^{\alpha_c} a_0^{1-\alpha_c}
%\ave{e_1^2}^{1-\alpha_c} m_{\rm c}^{1/3} Q_{\rm D}^*(m_{\rm
%c})^{\alpha_c-1}.
\label{eq:tcc} 
\end{eqnarray}
For 100\,km-sized planetesimals ($m_{\rm c} \approx 10^{22}$\,g) with
$\rms{e}_2 \sim 0.1$,
$\tau_{\rm cc} \la \tau_{\rm df}$. 
It should be noted that dynamical friction increases $\rms{e}_2$ during
the $\rms{e}_1$ damping, which shortens $\tau_{\rm cc}$ and elongates
$\tau_{\rm df}$ as discussed above. 
Therefore, even if $\tau_{\rm cc} \sim \tau_{\rm df}$ initially,  
$\tau_{\rm cc}$ eventually becomes much shorter than $\tau_{\rm df}$. 
%Larger planetesimals, expected to have a long $\tau_{\rm cc}$, are
%likely for the eccentricity damping of protoplanets without significant
%collisional fragmentation. 

Equations (\ref{eq:tdf}) and (\ref{eq:tcc}) show 
$\tau_{\rm df}/\tau_{\rm cc}$ is independent of $M_{\rm disk}$ and
$\Delta a$ but
depends on $\rms{e}_1$, $\rms{e}_2$, and $m_{\rm c}$. 
For $\rms{e}_1 = \rms{e}_2 = 0.03$, $P_{\rm DF} \approx 60$ so that 
$\tau_{\rm cc} \ll \tau_{\rm df} $ for $m_{\rm c}
= 10^{16}$\,g, while $\tau_{\rm df} \sim \tau_{\rm cc}$ for $m_{\rm c}
\sim 10^{19}$\,g. Therefore, as shown in
Figs.~\ref{fig:evo_ecc_inc_frag_mc5e-18_ecc0.03} and
\ref{fig:evo_ecc_inc_frag_mc5e-15_ecc0.03}, $e$ and $i$ of protoplanets insignificantly decrease
 for $m_{\rm c} = 10^{16}$\,g because of 
$\tau_{\rm cc} \ll \tau_{\rm df}$, while those are moderately damped for
$\tau_{\rm cc} \sim \tau_{\rm df}$ with $m_{\rm c} = 10^{19}$\,g. 
Therefore, the condition with $t_{\rm cc} \gg \tau_{\rm df}$ 
is required for the $e$ damping of protoplanets. 

Planets formed via giant impacts have high eccentricities and
inclinations. For Earth-sized planets formed in the giant impact stages,
the mean eccentricity is $\sim 0.1$ \citep{chambers01,kokubo06}, which
is much larger than the current eccentricities of Earth and Venus.  The
depletion timescale of a planetesimal disk with $m_{\rm c} \la
10^{21}\,$g is shorter than the $e$ damping timescale via dynamical
friction (see Eqs.~\ref{eq:tdf} and \ref{eq:tcc}).  Therefore, larger
planetesimals are required for $e$ damping of Earth-sized planets.  We
consider such large planetesimals are produced from a giant impact.  The
mass ratios of largest giant impact ejecta to parent protoplanets are $\sim
0.01$ \citep{genda15}.  Therefore, we consider $m_{\rm c} \sim
10^{26}$\,g, resulting in $\tau_{\rm cc} \gg \tau_{\rm df}$.

\begin{figure}[htb]
 \plotone{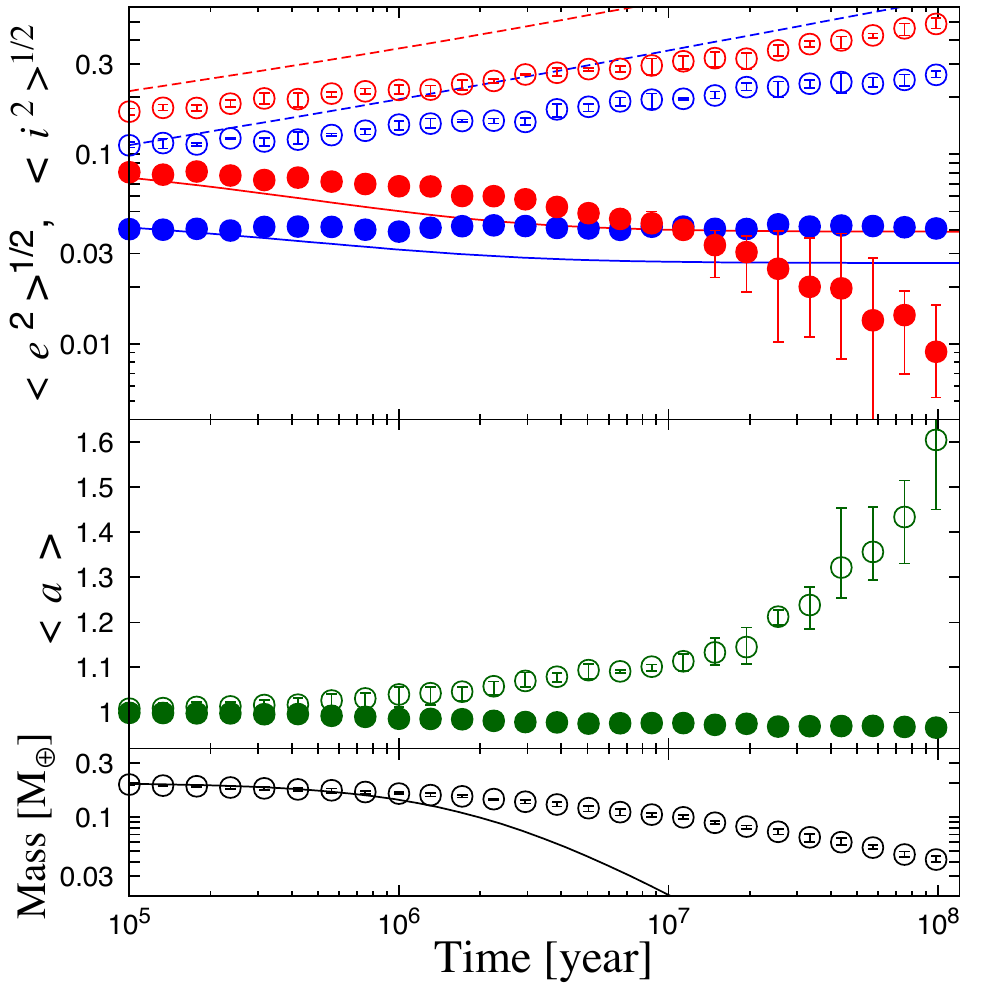} \figcaption{ The orbital
 and mass evolution of a planet and a swarm of giant impact ejecta. Top
 panel shows $\rms{e}$ (red) and $\rms{i}$ (blue) for the protoplanet
 (filled circles) and giant impact ejecta (open circles).  
Middle panel also shows $\ave{a}$ (green). 
Bottom panel
 represents the total mass of giant impact ejecta. We carry out three
 runs with different initial positions. The error bars indicate
 the minimum and maximum values. The lines are the same as
 Fig.~\ref{fig:evo_ecc_inc_frag_mc5e-18_ecc0.03}.  }
 \label{fig:evo_ecc_inc_frag_mc5e-8_ecc0.1}
\end{figure}

We perform the simulation for the orbital evolution of a
high-eccentricity planet with mass $M_\oplus$ in a swarm of giant impact
ejecta. \rev{Giant impact ejecta initially have similar orbits to parent
planets. However, $\varpi$ and
$\Omega$ are eventually distributed uniformly due to perturbation
from other planets.  The timescale to achieve a uniform distribution for
$\varpi$ and $\Omega$, which is roughly given by the precession rates for
$\varpi$ and $\Omega$ obtained from the secular perturbation theory 
\citep{Murray99}, is estimated as $\sim 10^5-10^6$\,years for
giant impact ejecta around 1AU perturbed by a Venus-like planet, which
is much shorter than the timescale for dynamical friction caused by
giant impact ejecta. Therefore, instead of ignoring perturbation from
other planets, we uniformly set $\varpi$ and $\Omega$ of giant impact
ejecta from the beginning.  } We initially set $a=1\,$AU and $e = 2i =
0.1$ for the planet and $a=0.95$--1.05\,AU and $\rms{e}=2\rms{i}=0.1$
for a giant-impact-ejecta disk composed of 120 super particles with
total mass $0.2M_\oplus$ and $m_{\rm c} = 10^{26}\,{\rm g}$
(Fig.~\ref{fig:evo_ecc_inc_frag_mc5e-8_ecc0.1}).  The evolutions of
$\rms{e}$ and $\rms{i}$ for giant impact ejecta and the planet
differs from those predicted by the analytical solutions given by the
time integration of Eqs.~(\ref{eq:dsigma_dt}), (\ref{eq:dedt}), and
(\ref{eq:didt}). Once $\rms{e}$ and/or $\rms{i}$ are much larger than
0.1, their orbits are controlled by the higher order terms for $\rms{e}$
and $\rms{i}$, which are ignored in Eqs.(\ref{eq:dedt}) and
(\ref{eq:didt}). The higher order terms make dynamical friction less
effective. Therefore, the variations of $\rms{e}$ and $\rms{i}$ for the
planet and planetesimals are less than those expected by the analytic
solution. However, $e$ of the planet becomes comparable to or larger
than the analytic
estimate until 10\,Myrs and decreases much greater than the analytic
estimate in several 10\,Myrs.  This is
caused by the orbital energy damping for the planet rather than its
angular momentum variation.  \rev{ If the angular momentum is fixed, the
energy damping given by $- \Delta a /a$ results in $\Delta e = (1-e^2)
\Delta a / 2 a e$, where $\Delta e$ and $\Delta a$ are the changes in
$e$ and $a$, respectively. Therefore, even a small energy damping of $\Delta a /a
\approx - 0.02$ leads to $\Delta e \approx - 0.1$ for the planet. } Giant impact ejecta
are scattered by the planet and tend to stay in the outer disk so that
the planet loses the orbital energy via the scatterings.  \rev{ This is
clearly shown in the semimajor-axis evolution for the planet and giant
impact ejecta (Fig.~\ref{fig:evo_ecc_inc_frag_mc5e-8_ecc0.1}). } The
collisional damping between the planet and giant impact ejecta also
reduces the orbital energy of the planet.  \rev{The accretion mass of
giant impact ejecta onto the planet is $0.086^{+0.004}_{-0.006}
M_\oplus$ in 10Myrs, $0.029_{-0.005}^{+0.011} M_\oplus$ for $t =
10$--30\,Myrs, and $0.029_{-0.004}^{+0.04} M_\oplus$ for $t =
30$--100\,Myrs. Although the $e$ damping is significant after 30\,Myrs,
the collisional damping expected from the accretion mass is slight.
Therefore, the energy loss for the planet due to the scattering of giant
impact ejecta mainly decreases its eccentricity.  }

The mass evolution of giant impact ejecta due to collisional
fragmentation is less important than that estimated by
Eq.~(\ref{eq:dsigma_dt}).  
\rev{Increase in $e$ of giant impact ejecta  due to planetary perturbation makes the giant-impact-ejecta
disk wider. 
%The original width of a giant-impact-ejecta
%disk becomes smaller than radial orbital widths $2 a e$ of single super
%particles.  
This decreases the surface density of giant impact ejecta, 
which reduces the efficiency of collisional fragmentation.  
In addition,
collisions between high-eccentricity giant impact ejecta are more
frequent around the planetary orbit. 
Collisional fragmentation between giant impact ejecta with similar $\varpi$
and $\Omega$ thus mainly occurs so that those relative velocities are
comparatively small (see Eqs.~\ref{eq:rel_e} and \ref{eq:rel_i}). The mass loss
by collisional fragmentation is thus less effective due to the increase of
ejecta eccentricity. Therefore,
the giant-impact-ejecta disk is maintained in a long timescale $\ga
10$\,Myr, resulting in the significant eccentricity damping for the
planet. 
}

\section{Discussion}
\label{sc:discussion}

In the giant impact stage, orbital instability of
protoplanets occurs in a timescale longer than 10\,Myrs 
\citep{chambers96,chambers98,iwasaki01,iwasaki02,kominami02,kokubo06}.
The protoplanets formed from the long-term evolution mainly have high
eccentricities ($\sim 0.1$) \citep{chambers01,kokubo06}, which are much
larger than the current eccentricities of Earth and Venus. A swarm of
planetesimals may damp the eccentricities of protoplanets due to
dynamical friction \citep{obrien06,raymond09,morishima10}. Instead,
dynamical friction increases eccentricities of planetesimals,
which induces collisional fragmentation between planetesimals.
The collisional cascade grinds planetesimals until small fragments are
brown out by radiation pressure, which results in the mass loss of the
planetesimal disk.  Collisional fragmentation may thus suppress the
eccentricity damping for protoplanets.  Therefore, we have investigated
the orbital interaction between protoplanets and planetesimals, taking
into account collisional fragmentation.

Eccentricities and inclinations of protoplanets are damped via dynamical
friction if we ignore collisional fragmentation (see
Fig. \ref{fig:orbital_dist}cd). However, collisional fragmentation
weakens dynamical friction (see Fig. \ref{fig:orbital_dist}ab). The mass
loss due to collisional fragmentation depends on the typical
planetesimal size of which planetesimals mainly determine the total
mass of the planetesimal disk.  Collisional depletion occurs in a short
timescale for planetesimal disks with small typical planetesimal sizes.
For 100km-sized or smaller planetesimals, the eccentricity damping is
ineffective due to collisional fragmentation
(Figs.~\ref{fig:evo_ecc_inc_frag_mc5e-18_ecc0.03} and
\ref{fig:evo_ecc_inc_frag_mc5e-15_ecc0.03}).

The primordial planetesimals may remain even in the giant impact stage.
Although the typical size of primordial planetesimals is not unknown,
the size may be on the order of 100\,km, similar to that of Main-Belt
asteroids \citep[e.g.,][]{kobayashi16}.  Due to collisional
fragmentation, dynamical friction for eccentricity damping is
ineffective in such a planetesimal disk.  On the other hand, the
increase in the eccentricities of protoplanets is required for the onset
of the orbital instability of protoplanets for giant impacts
\citep{iwasaki01}.  Although primordial planetesimal disks may not
explain the small eccentricities of Earth and Venus, such a disk does
not inhibit the onset and maintain of giant impact stages
\citep{walsh19}.  In addition, the direct formation of Earth and Venus
via the accretion of planetesimals is insignificant due to the depletion
of remnant planetesimal disks via collisional fragmentation
\citep{kobayashi+10,kobayashi13}.  Therefore, collisional fragmentation
supports the giant impact scenario to form Venus and Earth.

Giant impacts lead to the ejection of fragments as well as collisional
growth of protoplanets. The collisional ejecta from single impacts have
$0.1$--$0.3M_\oplus$, while the typical size of giant impact ejecta can
be as large as 1,000\,km \citep{genda15}.  The simulation for the
interaction between a protoplanet and giant impact ejecta shows
significant eccentricity damping of the protoplanet in $\ga 30$\,Myr
(see Fig.~\ref{fig:evo_ecc_inc_frag_mc5e-8_ecc0.1}).  The collisional
fragmentation is less effective in an originally narrow planetesimal
disk for high-eccentricity planetesimals, while the dynamical friction
is also less effective for high eccentricities and inclinations of
planetesimals. As a result, eccentricity damping for planets occurs in a
long timescale $\sim 100$\,Myr.

Giant impact ejecta are produced even in the early giant impact stage
\citep{genda15}. However, the orbital separations of protoplanets are so
narrow that giant impact ejecta are distributed widely due to
interactions with other protoplanets. In addition, the eccentricity
damping timescale due to dynamical friction is much longer than
collisional timescale between protoplanets. Therefore, sequent giant
impacts occurs due to insignificance of dynamical friction by giant
impact ejecta. On the other hand, stable orbital configurations of
protoplanets are achieved after tens of giant impacts. Such orbital
separations are wide enough to keep the orbital concentration of giant impact ejecta around
the orbits of single protoplanets. Therefore, dynamical friction by
giant impact fragments is effective in the late giant impact stage,
which may form low-eccentricity planets.

\rev{Finally we need to discuss the accuracy of the mass loss due to
collisional fragmentation.  As we discuss above, the collisional mass
loss mainly occurs after the collisional cascade is achieved. In the
collisional cascade, the mass loss is mainly determined by the total
ejecta mass from single collisions, and is insensitive to the mass
distribution of ejecta \citep{kobayashi10}. Therefore, the uncertainty
of collisional fragmentation mainly comes from $Q_{\rm D}^*$ at the
typical sizes of planetesimals (see Eq.~\ref{eq:tcc}). As discussed in
the previous studies \citep{kobayashi16,kobayashi18}, $Q_{\rm D}^*$ of
10\,km-sized or smaller primordial planetesimals may be much larger than
$Q_{\rm D}^*$ we set in the simulations. However, taking into account
the enhancement of $Q_{\rm D}^*$, 100\,km-sized or smaller primordial
planetesimals insignificantly work for the $e$ damping of
protoplanets. On the other hand, $Q_{\rm D}^*$ of 1000\,km-sized bodies
are mainly determined by the self-gravity of colliding bodies
\citep{kobayashi+10,kobayashi11,genda15b,genda17,suetsugu18}.  The
uncertainty is small for $Q_{\rm D}^*$ of largest ejecta of giant
impacts. Therefore, the $e$ damping of planets formed in the giant
impact stage is likely to be caused by giant impact ejecta.  }

\section{Summary}
\label{sc:summ}

Terrestrial planets are formed via giant impacts between Mars-sized
protoplanets. The resultant planets have larger eccentricities than the
current values for Earth and Venus. Dynamical friction with a
planetesimal disk is expected to damp the eccentricities of protoplanets. On the other
hand, the collisional cascade of planetesimals and blow-out of small
collisional fragments by radiation pressure decrease the
planetesimal-disk mass, which weakens dynamical friction with
planetesimals. 
%Therefore, 
%we have developed an N-body simulation code involving the mass loss due to
%the collisional cascade. 
%Using the code, we have investigated the orbital evolution of protoplanets in
%planetesimal disks, taking into account collisional fragmentation. 
Therefore, we have investigated the orbital evolution of planets with collisional fragmentation. Our findings are as follows. 

\begin{itemize}
 \item[1.] We have developed an N-body simulation code involving the
	   mass loss due to the collisional cascade. We have calculated
	   the mass evolution of a planetesimal disk composed of super
	   particles using the code, which reproduces the analytical
	   solution for the mass loss due to the collisional cascade
	   with high accuracy.
 \item[2.] We have investigated the evolution of orbits and masses of
	   protoplanets and planetesimals via gravitational interaction
	   and collisional fragmentation using the N-body code. If
	   collisional fragmentation is ignored, dynamical friction
	   damps eccentricities and inclinations for
	   protoplanets. However, collisional fragmentation suppresses
	   dynamical friction. The timescale ratio of dynamical friction
	   to collisional fragmentation depends on the typical
	   planetesimal size but not on the disk mass.  Even
	   a massive planetesimal disk composed of 100km-sized or
	   smaller planetesimals cannot damp eccentricities of
	   planets. \rev{Therefore, the disks composed of primordial
	   planetesimals are ineffective for the eccentricity damping for
	   planets. }
%	   On the other hand, if the typical planetesimal size
%	   is much larger, eccentricities of planets
%	   significantly decrease due to dynamical friction 
%	   even in a planetesimal disk with a mass smaller than
%	   planets. 
 \item[3.] Giant impacts eject collisional fragments. The total masses
	   of giant impact ejecta are several tenth of colliding
	   planets. The typical size of the largest ejecta is $\sim
	   1000$\,km. We have carried out simulations for a planet with
	   an Earth mass in
	   the disk with 0.2 Earth masses composed of giant impact
	   ejecta with largest ejecta $10^{26}$\,g.  Collisional mass
	   loss is insignificant for such a large typical size. 
	   Eccentricity of the
	   planet is damped from $0.1$ to $\sim 0.01$ due to
	   interactions with giant impact ejecta in $\ga
	   30$\,Myrs. Therefore, giant impact ejecta are possible to
	   decrease the eccentricities of planets formed via giant
	   impacts, comparable to those of Earth and Venus.
\end{itemize}

\rev{
We thank the reviewer for beneficial comments.}
The work is supported by
Grants-in-Aid for Scientific Research
(17K05632, 17H01105, 17H01103, 18H05436, 18H05438)
from MEXT of Japan and by JSPS Core-to-Core Program ``International Network of Planetary Sciences''. 

%\appendix

\end{document}